\documentclass[aps,prd,twocolumn,superscriptaddress,showpacs,amsmath,amssymb,nofootinbib,eqsecnum, preprintnumbers]{revtex4-1}
\usepackage{graphicx}
\usepackage{epstopdf}
\usepackage{float}
\usepackage{color}
\usepackage{tensor}
\usepackage[toc,page]{appendix}
\usepackage[usenames,dvipsnames]{xcolor}

\newcommand{\be}{\begin{equation}}
\newcommand{\ee}{\end{equation}}
\newcommand{\ba}{\begin{eqnarray}}
\newcommand{\ea}{\end{eqnarray}}

\newcommand{\beq}{\begin{equation}}
\newcommand{\eeq}{\end{equation}}
\newcommand{\beqa}{\begin{eqnarray}}
\newcommand{\eeqa}{\end{eqnarray}}
\newcommand{\nn}{\nonumber}

\newcommand{\sccur}{{\mathcal{R}}}

\begin{document}

\title{Reentrant phase transitions and van der Waals behaviour for hairy black holes}

\author{Robie A. Hennigar}
\email{rhennigar@uwaterloo.ca}
\affiliation{Department of Physics and Astronomy, University of Waterloo,
Waterloo, Ontario, Canada, N2L 3G1}

\author{Robert B. Mann}
\email{rbmann@sciborg.uwaterloo.ca}
\affiliation{Department of Physics and Astronomy, University of Waterloo,
Waterloo, Ontario, Canada, N2L 3G1}

\begin{abstract}
We study the extended phase space thermodynamics for hairy AdS black hole solutions to Einstein-Maxwell-$\Lambda$ theory conformally coupled to a scalar field in five dimensions.  We find these solutions to exhibit van der Waals behaviour in both the charged/uncharged cases, and reentrant phase transitions in the charged case.  This is the first example of reentrant phase transitions in a five dimensional gravitational system which does not include higher curvature corrections. 
\end{abstract}


\maketitle


\section{Introduction}

For over forty years the study of black hole thermodynamics has remained an active field of inquiry.  Of particular interest are the thermodynamics of Anti de Sitter (AdS) black holes which, through proposed gauge/gravity dualities, represent the holographic duals of strongly coupled conformal field theories, and for which thermodynamic equilibrium is straightforward to define. 

Recently there has been flourishing interest in the subject of {\it extended phase space thermodynamics}, where the cosmological constant $\Lambda$ is treated as a thermodynamic variable~\cite{Creighton:1995au} analogous to pressure \cite{CaldarelliEtal:2000, Dolan:2010ha, Dolan:2011xt, Dolan:2014jva}.  In this paradigm, the mass of the black hole is identified as the enthalpy of the spacetime \cite{Kastor:2009wy} which leads to a remarkable correspondence between the thermal physics of black holes and simple everyday substances \cite{Kubiznak:2014zwa}.  In initial work, it was shown that there exists a complete physical analogy between the critical behaviour of the $4$-dimensional Reissner-Nordstrom AdS black hole and the van der Waals liquid-gas system \cite{Kubiznak:2012wp}.  Later investigations found examples of  {\it triple points}, analogues of the {\it solid/liquid/gas} transition and {\it reentrant phase transitions} for $6$-dimensional, doubly-rotating Kerr-AdS black holes \cite{Altamirano:2013uqa, Altamirano:2013ane, Altamirano:2014tva}.  Extensions of this work to higher curvature theories of gravity have yielded more exotic results, such a {\it multiple reentrant phase transitions} and {\it isolated critical points} \cite{Frassino:2014pha, Hennigar:2015esa, Dolan:2014vba}.  Numerous studies have confirmed similar results for a broad range of black hole solutions \cite{Gunasekaran:2012dq, Cai:2013qga, Hendi:2012um, Mo:2014mba}, while other studies have used this black hole chemistry framework \cite{Kubiznak:2014zwa} to develop ideas such as thermodynamically inspired black holes \cite{Rajagopal:2014ewa, Delsate:2014zma, Setare:2015xaa}, entropy inequalities \cite{Cvetic:2010jb, Hennigar:2014cfa}, and the notion of a holographic heat engine \cite{Johnson:2014yja}.

The proposed relationship between the cosmological constant and the thermodynamic pressure is
\be 
P = -\frac{\Lambda}{8 \pi} = \frac{(d-1)(d-2)}{16 \pi l^2} \, ,
\ee
where $d$ is the number of spacetime dimensions.  The quantity conjugate to $P$ is the {\it thermodynamic volume}, $V$, which is defined so that the extended first law of black hole thermodynamics holds:
\be 
\delta M = T \delta S + \sum_i \Omega_i \delta J_i + \Phi \delta Q + V \delta P \, .
\ee
Here $M$ is the mass, $T$ is the temperature, $S$ is the entropy, $Q$ is the electric charge, and $J_i$ are the angular momenta of the black hole.  The quantities $\Phi$ and $\Omega_i$ are the electric potential and angular velocities of the horizon, which are measured relative to infinity.  An Eulerian scaling argument can be applied to the first law to derive the Smarr formula,
\be 
\frac{d-3}{d-2}M = TS + \sum_i \Omega_i J_i + \frac{d-3}{d-2} \Phi Q - \frac{2}{d-2} VP 
\ee
which is also the result of   geometric arguments based on the Komar formula \cite{Kastor:2009wy}.

Here we begin the first exploration of hairy black AdS hole solutions of Einstein-Maxwell-$\Lambda$ theory conformally coupled to a scalar field in the context of extended phase space thermodynamics.  While there are well-established no hair theorems for four dimensional general relativity conformally coupled to a scalar field theory in asymptotically flat spacetime, the situation is more interesting  in asymptotically (locally) AdS space where conformally dressed solutions have long been known to exist \cite{Martinez:1996gn}.  The AdS solutions are inherently interesting since, through the AdS/CFT correspondence, the condensation of a scalar field around a $4$-dimensional AdS black hole provides a holographic model of superconductors \cite{Hartnoll:2008vx, Gubser:2008px}. The 5-dimensional theory we consider here was originally developed and studied in \cite{Oliva:2011np, Galante:2015voa, Giribet:2014bva, Giribet:2014fla} and has the significant advantage of being a simple and tractable model for studying  phase transitions of hairy black holes that includes the back-reaction of the scalar field on the metric.  

In what follows we first write explicitly the solution of \cite{Galante:2015voa} for the case where the constant $(t,r)$ hypersurface is a manifold of constant postive, zero, or negative curvature and calculate the thermodynamic quantities. While no interesting results are obtained in the flat or hyperbolic cases, we show that these solutions exhibit van der Waals behaviour and can undergo reentrant phase transitions in the spherical case.  These are the first examples of reentrant phase transitions for five dimensional black holes outside of higher curvature gravity theories.  In an appendix we address the negative mass solutions of this theory (which can have up to four horizons) and discuss their extremal limits.

\section{Solution \& Thermodynamics}

The action for the theory is given by \cite{Galante:2015voa, Giribet:2014bva, Giribet:2014fla},
\be\label{theory} 
{\cal I} =  \frac{1}{\kappa} \int d^5 x \sqrt{-g} \left(R - 2 \Lambda - \frac{1}{4}F^2 + \kappa {\cal L}_m(\phi, \nabla \phi) \right)
\ee
where
\ba 
{\cal L}_m &=& b_0 \phi^{15} + b_1 \phi^7 \tensor{S}{_\mu_\nu^\mu^\nu} + b_2 \phi^{-1}(\tensor{S}{_\mu_\gamma^\mu^\gamma}\tensor{S}{_\nu_\delta^\nu^\delta} 
\nn\\
&&- 4 \tensor{S}{_\mu_\gamma^\nu^\gamma}
\tensor{S}{_\nu_\delta^\mu^\delta} + \tensor{S}{_\mu_\nu^\gamma^\delta}\tensor{S}{^\nu^\mu_\gamma_\delta})
\ea
with $b_0, b_1$ and $b_2$ the coupling constants of the conformal field theory, and
\ba 
\tensor{S}{_\mu_\nu^\gamma^\delta} &=& \phi^2 \tensor{R}{_\mu_\nu^\gamma^\delta} - 12 \delta^{[\gamma}_{[\mu} \delta^{\delta]}_{\nu]}\nabla_\rho\phi\nabla^\rho\phi 
\nn\\
&&- 48 \phi \delta^{[\gamma}_{[\mu} \nabla_{\nu]} \nabla^{\delta]} \phi + 18 \delta^{[\gamma}_{[\mu}\nabla_{\nu]}\phi\nabla^{\delta]}\phi \,.
\ea

The five-dimensional solution to the theory \eqref{theory} is given by
\be\label{metric} 
ds^2 = -f dt^2 + \frac{dr^2}{f} + r^2 d\Omega^2_{(k)3}
\ee
where 
\be 
f = k-\frac{m}{r^2} - \frac{q}{r^3} + \frac{e^2}{r^4} + \frac{r^2}{l^2}
\ee
and $d\Omega^2_{(k)3}$ is the line element on a three-dimensional surface of constant positive, zero, or negative curvature ($k=1,0$ or $-1$, respectively).  Here $e$ represents the electric charge, $m$ the mass parameter of the solution, and $q$ is given in terms of the coupling constants of the scalar field,
\be 
q = \frac{64 \pi}{5} \varepsilon k b_1 \left(\frac{-18 k b_1 }{5b_0} \right)^{3/2} \,,
\ee
with $\varepsilon$ taking the values $\varepsilon=-1,0,1$. Consequently precisely specifying the coupling constants of the scalar field does not uniquely determine the value of $q$ appearing in the solution; instead
$q$ is a discrete parameter taking values $q=0, \pm |q|$. In order to satisfy the field equations, the scalar coupling constants must obey the constraint
\be 
10b_0b_2 = 9 b_1^2 \, .
\ee
Note also that for planar solutions, i.e. $k=0$, we have $q=0$. In other words, in the planar case, there is no hair.  Therefore, in the following we shall  consider only the $k= \pm 1$ cases.

The hair parameter  $q$ is not a conserved charge corresponding to some symmetry---it can change provided the scalar field coupling constants are dynamic.   Here, in developing the first law and Smarr relation, we shall consider all couplings to be dynamical.  In particular this means that we shall consider $q$ to be a continuous, real parameter. 
The Maxwell potential is given by,
\be 
A_t = \frac{\sqrt{3} e}{r^2}
\ee
and the field strength follows in the standard way: $F_{\mu \nu}~=~ \partial_\mu A_\nu - \partial_\nu A_\mu$, while the scalar field configuration is given by (for $ k= \pm 1$),
\be 
\phi(r) = \frac{n}{r^{1/3}} \, , \quad n = \varepsilon \left(- \frac{18 k} {10} \frac{ b_1}{b_0} \right)^{1/6}
\ee

Note that the metric \eqref{metric} admits zero and negative mass solutions for $k = \pm 1$ for $q >0$. These negative mass solutions were regarded as pathological in~\cite{Galante:2015voa, Giribet:2014fla}.  In the body of this paper we consider only positive mass black holes, and perform a short study of these more exotic possibilities in the appendix.

A number of the thermodynamic properties of this solution were discussed in \cite{Galante:2015voa} for the case of $k=1$.  Here we have generalized the calculations to include the $k=-1$ case.  The thermodynamic quantities are given by,
\ba 
M &=& \frac{3 \omega_{3(k)} }{16 \pi }  m, \, \, \, Q = - \frac{\omega_{3(k)}  \sqrt{3}}{16 \pi} e , \, \, \,  S = \omega_{3(k)} \left(\frac{r_+^3}{4} - \frac{5}{8} q \right),
\nn\\
T &=& \frac{1}{\pi l^2 r_+^4} \left[-\frac{e^2l^2}{2r_+} + \frac{ql^2}{4} + \frac{ k l^2r_+^3}{2} + r_+^5  \right] ,
\ea
where $\omega_{3(k)}$ is the volume of the compact $3$-dimensional manifold with line element $d\Omega^2_{(k)3}$. The mass, $M$, and electric charge, $Q$, were calculated in \cite{Galante:2015voa} using the Regge-Teitelboim approach and we have arrived at the same results for general $k$. The temperature was calculated by requiring the absence of conical singularities in the Euclidean section. We verified the entropy does not change for the case of $k=-1$ by computing the Wald entropy \cite{Wald:1993nt}.   Note that, for positive values of $q$, the entropy can be negative.  We shall regard black holes with negative entropy as unphysical in our studies. 

The thermodynamic quantities above satisfy the extended first law,
\be 
dM = T dS + VdP + \Phi dQ + K dq
\ee
provided that
\ba 
V &=& \frac{\omega_{3(k)}}{4} r_+^4\, , \quad \Phi = -\frac{2\sqrt{3}}{r_+^2}e\,,
\\
K &=& \frac{\omega_{3(k)}}{32 l^2 r_+^5} \left(20r_+^6 +2r_+^4 l^2 (5k - 3) + 5l^2qr_+ - 10e^2l^2 \right) \,. \nn
\ea
The Smarr formula consistent with scaling,
\be 
2M = 3TS - 2VP + 2\Phi Q + 3 q K \, ,
\ee
is satisfied as well.

To determine the equation of state we simply solve the expression for temperature in terms of the pressure.  We identify the specific volume as $v = 4r_+/3$ both so that the thermodynamic quantities have the correct physical dimensions, and for the convenience of this expression compared to the thermodynamic volume in contextualizing this work in the existing literature.  In a number of cases, the specific volume can be understood as the ratio of the thermodynamic volume to the number of degrees of freedom of the system, $\tilde{v} = V/N$ where, for a black hole of area $A$,
\be 
N = \frac{1}{4} \frac{d-2}{d-1} \frac{A}{\ell_P^2}
\ee  
and $\ell_P$ is the Planck length (cf. footnote 5 of \cite{Altamirano:2014tva}). However, in some cases like the one at hand this formula fails to be appropriate.\footnote{Other cases include rotating black holes and black holes in higher curvature gravity.} The entropy is no longer simply one quarter of the horizon area and therefore $A$ does not represent the degrees of freedom in the system.  Modifying the definition by replacing $A$ with $S$ would also not be appropriate here, since then the specific and thermodynamic volumes would not be related in a natural way.  For these reasons, we utilize the specific volume in the form $v= 4r_+/3$, which is consistent with the dimensional arguments in~\cite{Gunasekaran:2012dq}.  

Of course, we must be careful in using the specific volume since certain quantities depend on the choice of volume in a meaningful way---for example, the result of integrating the equal area law~\cite{Wei:2014qwa}. However, in the analysis we perform here such problems do not arise, and our results are qualitatively no different from what would be obtained had we used the thermodynamic volume  explicitly.  These comments aside, we find the equation of state to be,
\be 
P = \frac{T}{v} - \frac{2 k}{3 \pi v^2} + \frac{512}{243}\frac{e^2}{\pi v^6} - \frac{64}{81}\frac{q}{\pi v^5} \, .
\ee 
The Gibbs free energy for our system is given by
\ba 
G &=& M - TS
\nn\\
&=& \omega_{3(k)} \left[ \frac{9 k v^2}{256 \pi} - \frac{27 v^4 P}{1024} + \frac{40  q^2}{81 \pi  v^4}  \right.
\\
&&+ \left. \left(\frac{5  P v}{8} + \frac{5k - 4}{12 \pi v} \right)q + \left(\frac{5}{9 \pi v^2} - \frac{320  q}{243 \pi v^5} \right) e^2 \right] \nn \,.
\ea
We are now ready to explore the phase structure of these hairy black hole solutions.

\section{Critical Behaviour}

Here we are interested in studying the criticality and phase structure of these solutions.  We first note that in the case $q=0$ the solution reduces to the $5$-dimensional Reisner-Nordstrom-AdS black hole solution, which was carefully studied in \cite{Gunasekaran:2012dq}.  Therefore we will exclusively focus on cases of non-zero $q$, with a particular interest in seeing if this ``hairy" parameter leads to any significant changes in the thermodynamic behaviour.  We dedicate the following two subsections to the exploration of the thermodynamic behaviour for spherical black holes with $k=1$ and hyperbolic black holes with $k=-1$.  Following this, we specialize to the case where the entropy is zero at the critical point. 

\subsection{Spherical}
\subsubsection{Uncharged case}

We begin by studying the behaviour of these black holes in the absence of a Maxwell field, i.e. $e=0$.  We find that the equation of state admits a single critical point for $q <0$ with the critical values,
\ba 
T_c &=& - \frac{3}{20} \frac{(-5q)^{2/3}}{\pi q}\,,\quad v_c = \frac{4}{3} (-5q)^{1/3} \, ,
\nn\\
P_c &=& \frac{9}{200 \pi} \left(-\frac{\sqrt{5}}{q} \right)^{2/3} \,.
\ea
The ratio of the critical values is
\be 
\frac{P_cv_c}{T_c} = \frac{2}{5}
\ee
which is different from both the $5$-dimensional RN-AdS solution and the van der Waals fluid. Expanding the equation of state about the critical point in terms of the parameters $\omega = v/v_c-1$ and $\tau = T/T_c-1$ we find,
\be 
\frac{P}{P_c} = 1 + \frac{5}{2}\tau - \frac{5}{2} \omega \tau - \frac{5}{3} \omega^3  + {\cal O}(\tau\omega^2, \omega^4)
\ee
thus, from the analysis in \cite{Gunasekaran:2012dq} we conclude that the critical point is characterized by the standard mean field theory exponents,
\be 
\alpha = 0 \, , \quad \beta = \frac{1}{2} \,,\quad \gamma = 1\, , \quad \delta = 3\,.
\ee

\begin{figure*}[htp]
\includegraphics[width=0.3\textwidth]{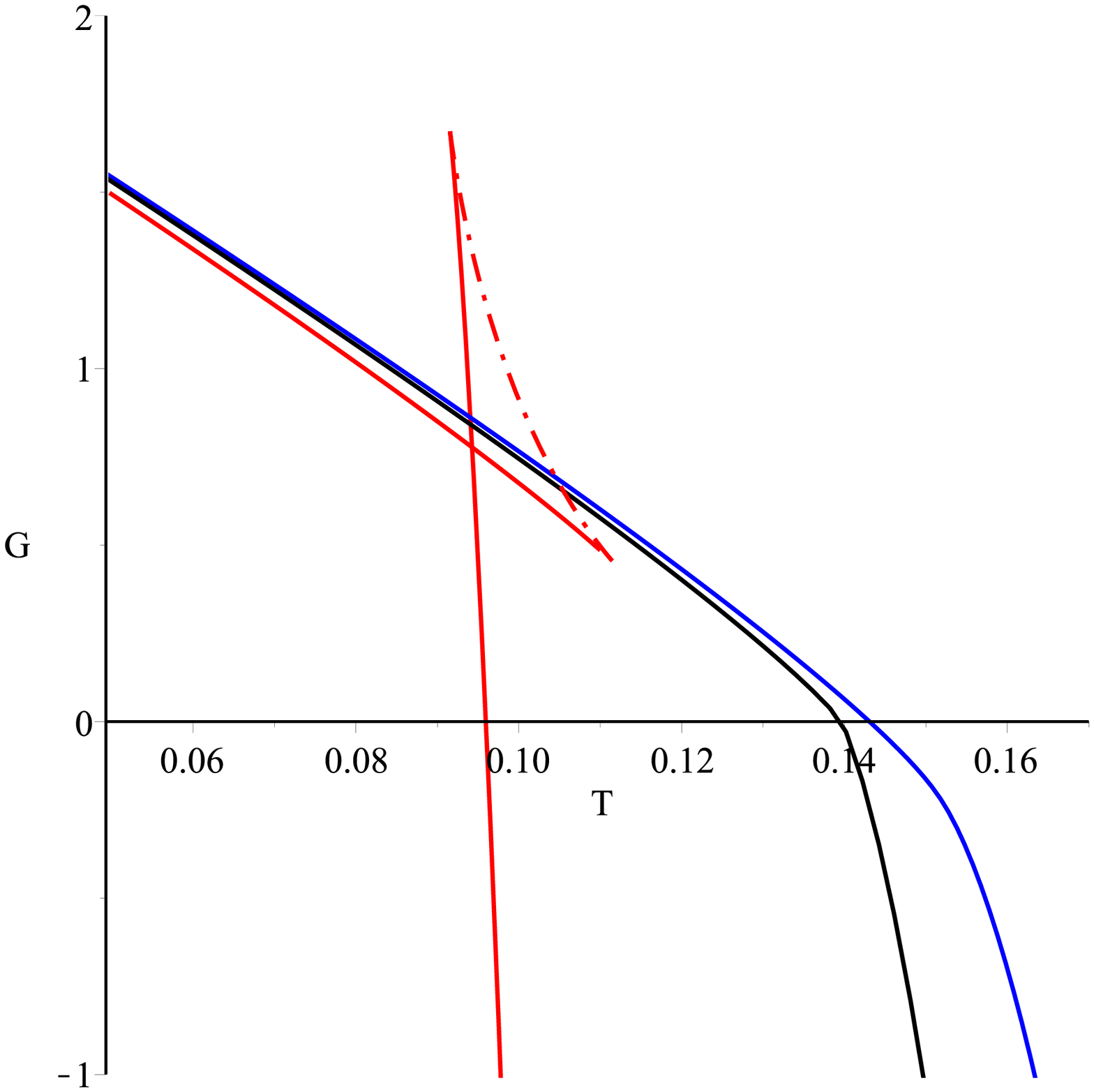}
\includegraphics[width=0.3\textwidth]{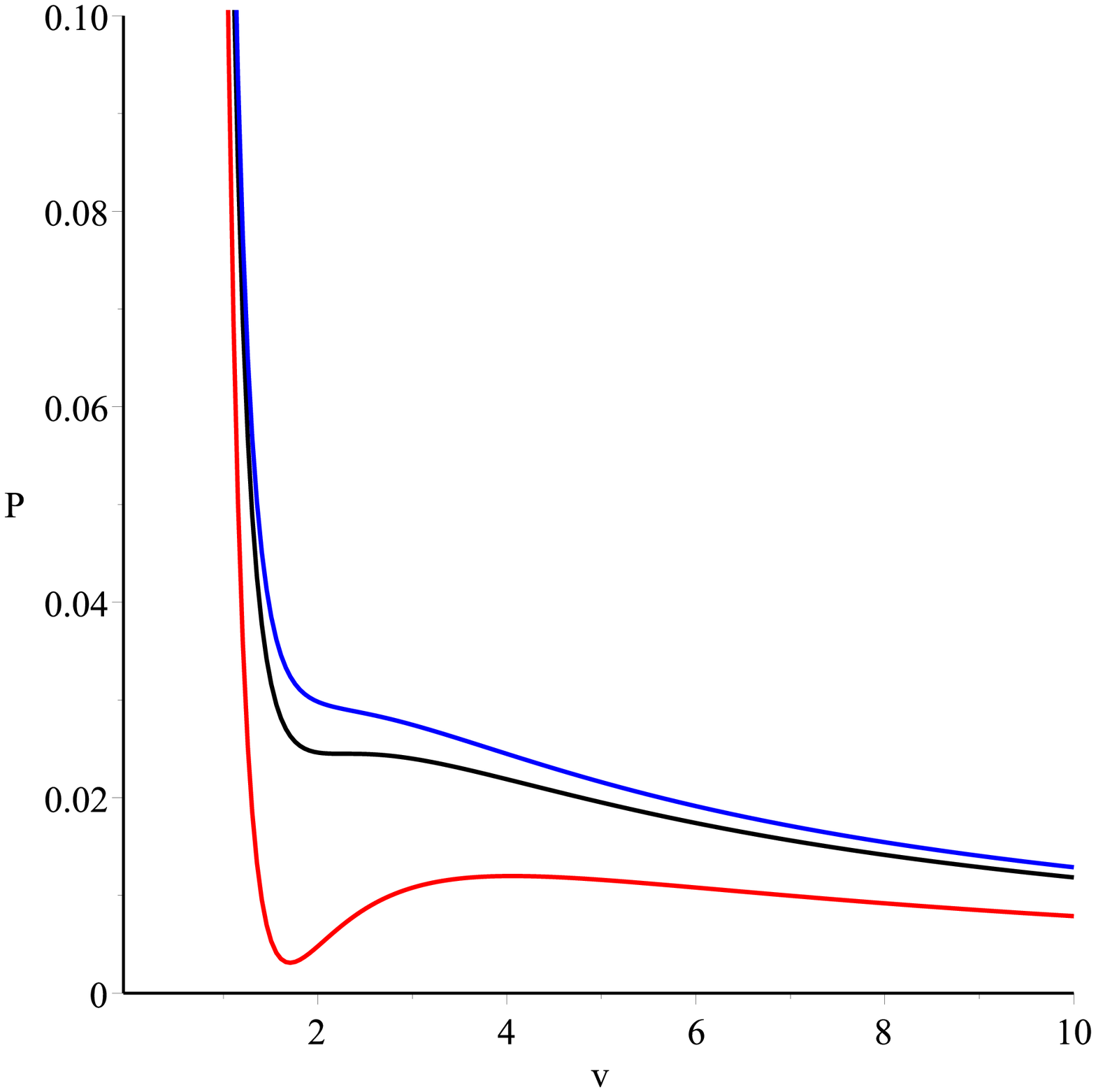}
\includegraphics[width=0.3\textwidth]{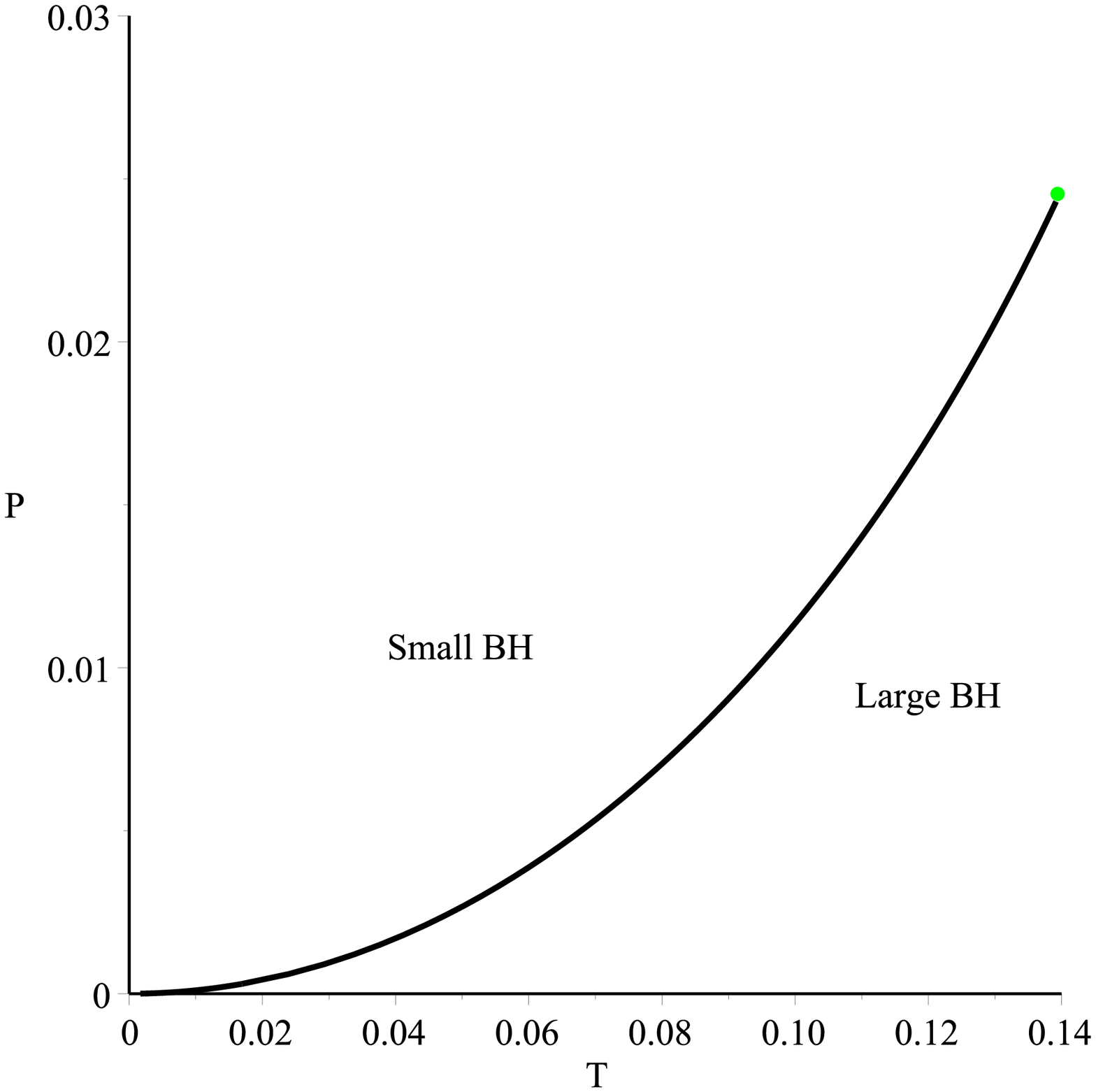}
\caption{{\bf Uncharged case: van der Waals behaviour}: $k=1, q=-1$. {\it Left}: $G$ vs. $T$ plot with the red, black and blue lines showing $P < P_c$, $P=P_c$ and $P>P_c$, respectively.  Below the critical pressure, we observe classic swallowtail behaviour.   {\it Center}: $P$ vs. $v$ plot plot with the red, black and blue lines showing $T < T_c$, $T=T_c$ and $T>T_c$, respectively.  For $T < T_c$ we observe a van der Waal-type oscillation.  {\it Right}: $P$ vs. $T$ coexistence plot.  Here we observe standard van der Waals behaviour, with the coexistence line beginning at the origin and terminating at the critical point (shown here as a green dot).}
\label{uncharged_plots}
\end{figure*}

The critical behaviour in the uncharged case is displayed in Figure~\ref{uncharged_plots} for the specific case $q=-1$, which highlights the salient features of the $q <0$ case.  The leftmost plot illustrates the typical behaviour of the Gibbs free energy.  We see that for $P < P_c$ the Gibbs energy displays the characteristic swallowtail shape associated with the first order phase transition.  The center plot shows representative $P-v$ curves, in which, for $T < T_c$ we see the van der Waals oscillation.  The rightmost figure shows a coexistence plot in $(P,T)$ space.  The black line marks the boundary which separates the two distinct phases---here, small and large black holes.  The coexistence line begins at the origin and terminates at the critical point, which is marked in this plot by the green dot.  This is standard van der Waals behaviour.

\begin{figure}[htp]
\centering
\includegraphics[width=0.3\textwidth]{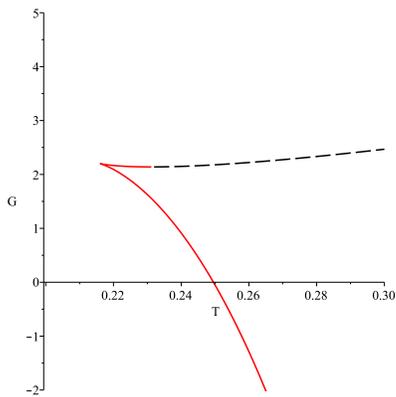}
\caption{{\bf Uncharged case: Gibbs free energy}: $k=1, q=1$.  A representative $G$ vs. $T$ plot for $P=0.05$ and $q=1$.  The red lines represent physical branches of the Gibbs energy, while the dashed black line corresponds to negative entropy black holes. The Gibbs free energy displays a cusp, the upper branch of which terminates at finite temperature due to enforcing positivity of entropy.  At temperatures below the cusp, no black hole solutions exist.}
\label{uncharged_cusp}
\end{figure}

As mentioned above, there is no critical point when $q \ge 0$. However, we must still consider this case explicitly since there could be zeroth order phase transitions induced by positivity of entropy considerations.  Transitions of this type were found in previous investigations focusing on higher curvature gravity \cite{Frassino:2014pha, Hennigar:2015esa} where enforcing positivity of entropy resulted in discontinuities in the Gibbs free energy, forcing the system to ``jump" from one minimal branch to another.  For positive $q$ we find that the Gibbs free energy exhibits a cusp, and while enforcing positivity of entropy introduces discontinuities, they always occur in the upper branch, as shown in Figure~\ref{uncharged_cusp}.  Therefore, enforcing positivity of entropy in the $q >0$ case leads to no interesting behaviour in the uncharged case.

\subsubsection{Charged case}

In this section we expand our analysis to include a non-vanishing charge parameter, $e$.  This complicates the analysis from an analytical point of view, as the expressions for the critical points no longer take a simple form.  We proceed numerically where appropriate.

\begin{figure}[htp]
\centering
\includegraphics[width=0.3\textwidth]{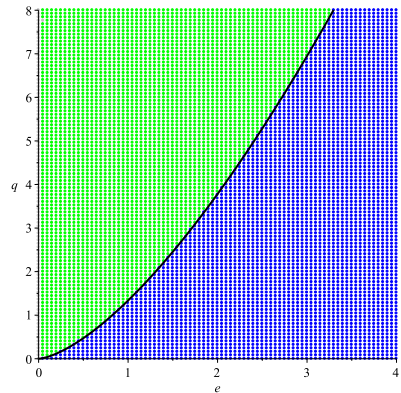}
\caption{{\bf Charged case: $q$-$e$ parameter space}: $k=1$.  A plot of a region in $q$-$e$ space showing the number of physical critical points.  Blue dots indicate a physical critical point having positive $(v_c, T_c, P_c)$ and positive entropy.  Green dots indicate unphysical critical points: the critical values are positive, but the entropy is negative at the would-be critical point.  The solid black line maps out the zero entropy boundary given by $q \approx 1.3375 e^{3/2}$.  For this value of $q$, the critical point corresponds to a zero entropy black hole.}
\label{q_e_scrape}
\end{figure}

When $q<0$ there is a single critical point present for any value of $e$.  Further, the positivity of entropy is trivially satisfied for this case.  No new features emerge here and Figure~\ref{uncharged_plots} is sufficient to describe the physics at play---there are no qualitative differences.  We observe a first order phase transition with classic swallowtail behaviour for the Gibbs free energy.  The $P-v$ and $P-T$ plots display standard van der Waals behaviour.  The critical points are characterized by the mean field theory exponents.

\begin{figure*}[htp]
\centering
\includegraphics[width=0.3\textwidth]{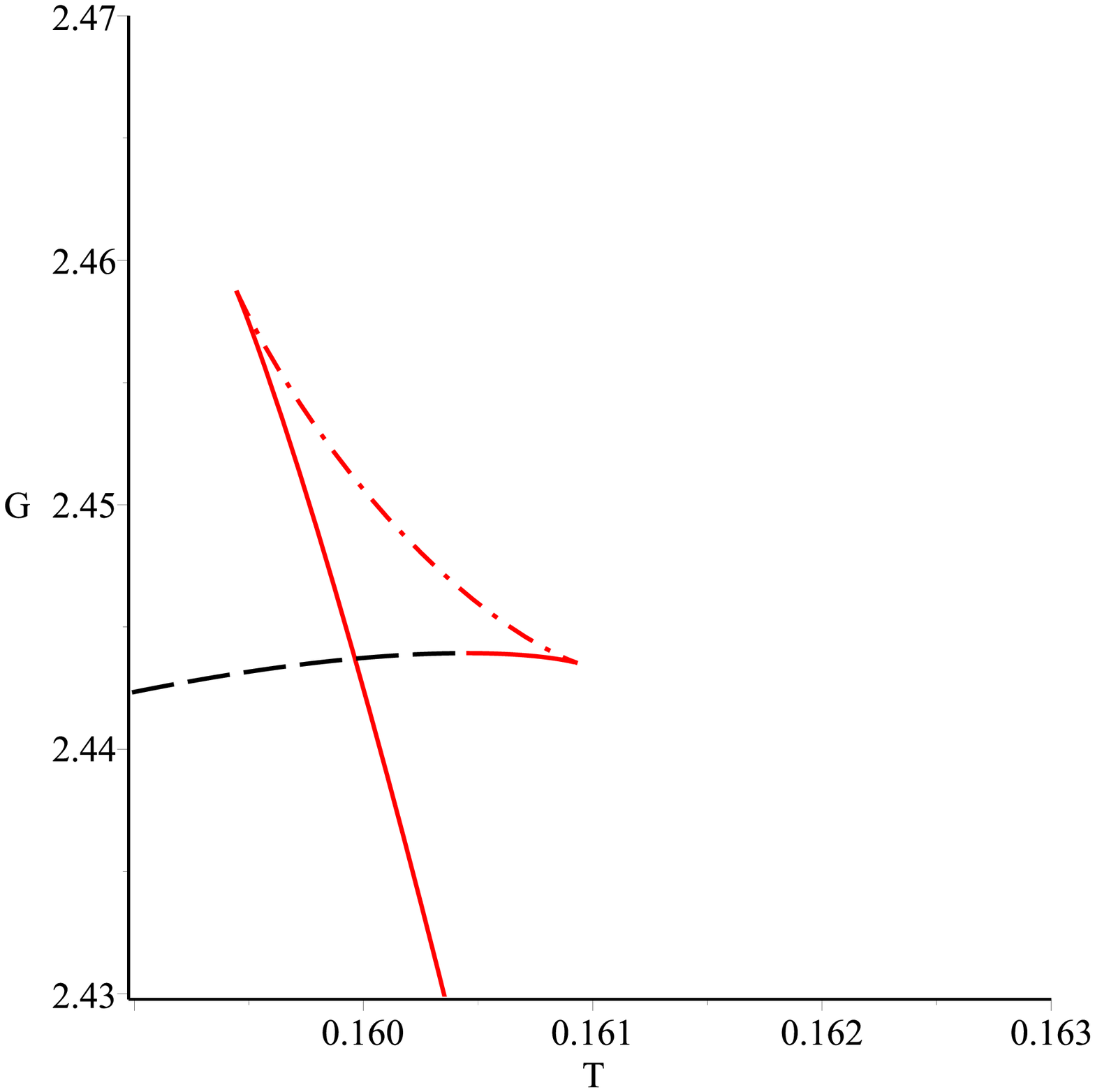}
\includegraphics[width=0.3\textwidth]{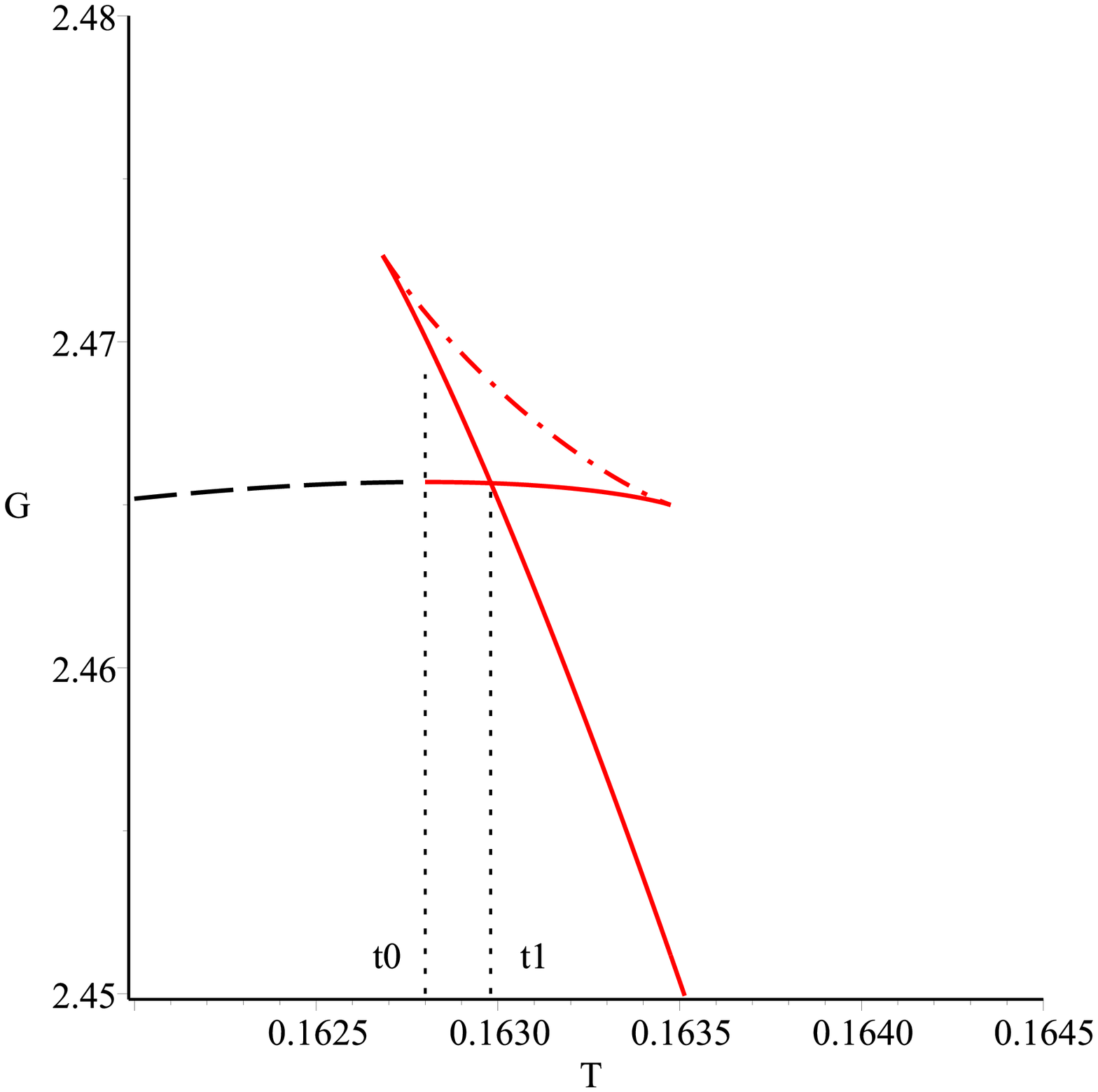}
\includegraphics[width=0.3\textwidth]{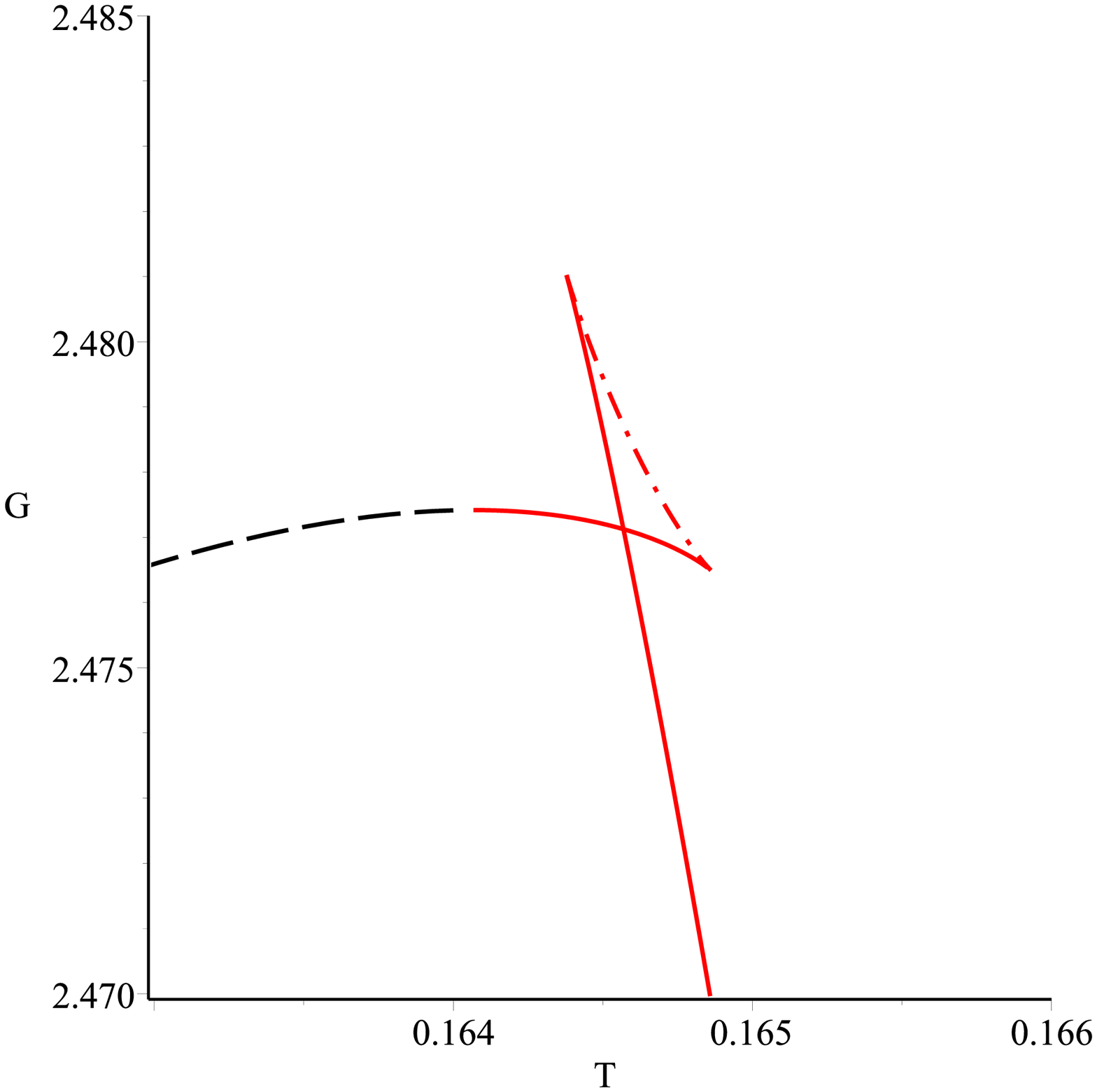}
\caption{{\bf Reentrant phase transition: Gibbs free energy}: $k=1$, $e=q=1$. These plots show the Gibbs free energy for $P=0.031, 0.0313$, and $ 0.032$ (left to right).  Due to positivity of entropy constraints, one branch becomes unphysical below a certain temperature, which changes depending on the pressure---in the plots, this is shown with the dashed black curves. For a certain window of pressures we see the behaviour exemplified by the center figure.  Such a configuration leads to a reentrant phase transition, since a monotonic variation of the temperature leads to a zeroth-order large/small black hole transition at $t_0$, followed by a first order small/large black hole transition at $t_1$.   }
\label{gibbs_reentrant}
\end{figure*}

\begin{figure*}[htp]
\centering
\includegraphics[width=0.33\textwidth]{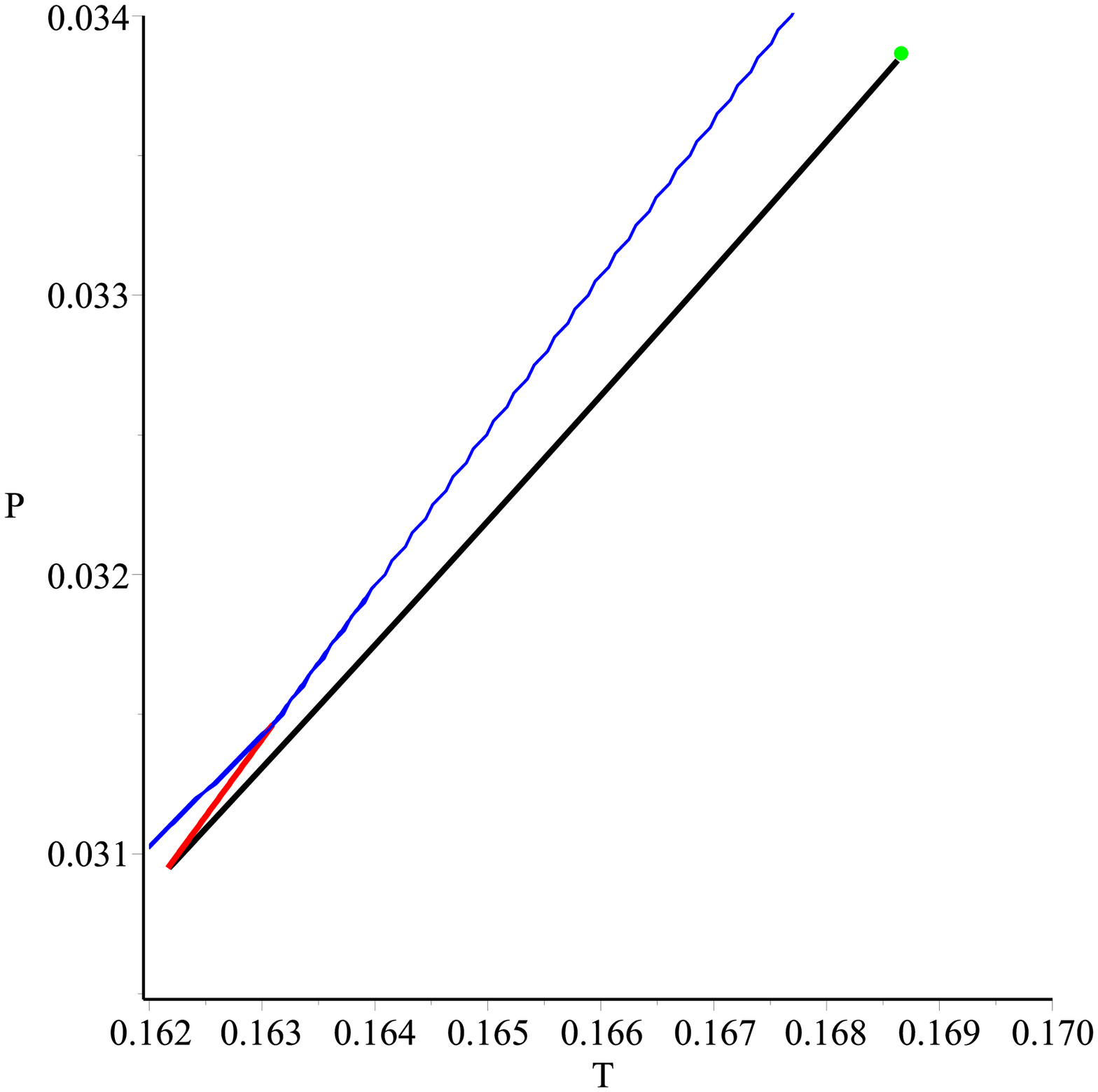} 
\quad
\includegraphics[width=0.33\textwidth]{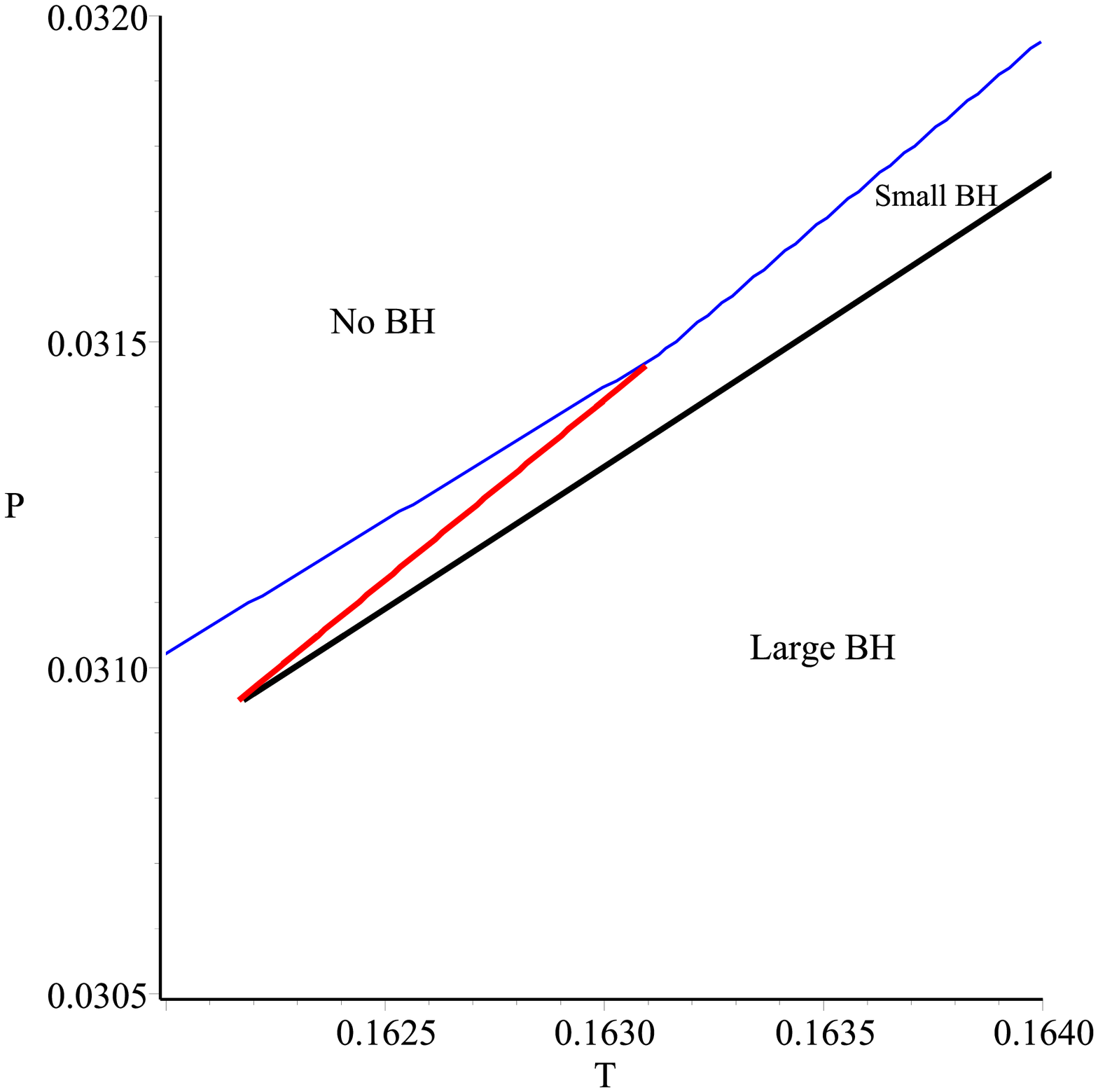}
\caption{{\bf Reentrant phase transition: coexistence plots}: $k=1$, $e=q=1$. {\it Left}: $P-T$ coexistence plot showing zeroth and first order phase transitions (red and black curves, respectively).  To the left of the blue line the black holes have negative entropy, and we therefore regard them as unphysical. {\it Right}: A zoomed-in copy of the left plot highlighting the presence of a large/small/large black hole reentrant phase transition for a small range of pressures.}
\label{coexistence_reentrant}
\end{figure*}

For $q>0$, the equation of state admits a single inflection point, but whether this inflection point is a physical critical point depends on the positivity of entropy conditions.  To illustrate this, consider Figure~\ref{q_e_scrape}, which maps out the number of critical points for positive $q$.  This figure shows clearly that, for some $(q,e)$ combinations the would-be critical point occurs for parameter values which correspond to a negative entropy black hole.  In fact, we can understand from this plot that, for a given charge $e$, there will be some threshold value $q_0$ such that for all $q > q_0$ the critical point occurs for a negative entropy solution.  By performing an analysis identical to that of the previous section we confirm that the critical points are characterized by the mean field theory exponents. 

In regions of $(q,e)$ space where the critical point is unphysical, we find the Gibbs free energy has a cusp qualitatively identical to Figure~\ref{uncharged_cusp}.  We have found no examples of phase transitions of any kind in these regions. 

To understand the thermodynamic behaviour of these solutions when the critical point is physical we discuss the $q=e=1$ case, which provides a representative analysis.  For this case, enforcing the positivity of entropy condition requires us to regard the Gibbs free energy as unphysical in certain parameter regions.  In contrast to the uncharged case, this leads to new and interesting behaviour: in addition to standard van der Waals behaviour we observe instances of reentrant phase transitions over small ranges of pressure.  These phase transitions, first observed in nicotine/water mixtures, are so-named because a monotonic variation of one thermodynamic parameter leads, after two or more phase transitions, to a final state which is macroscopically identical to the initial state \cite{hudson1904mutual}.   

The structure of the Gibbs free energy which leads to this behaviour is portrayed in Figure~\ref{gibbs_reentrant}.  Each plot in this figure corresponds to a different value of pressure which  increases going from left to right, starting at $P=0.031$ in the leftmost plot and ending at $P=0.032$ in the rightmost.  In each plot, below a certain temperature one branch of the Gibbs free energy (the one which is closest to being horizontal) is unphysical due to the positive entropy condition.  As the pressure increases, this branch becomes closer to, and eventually intersects, the minimal branch of the Gibbs energy.  When this intersection happens we have, for a narrow window of pressures, a reentrant phase transition, as shown in the center plot for $P=0.0313$.  A system described by the center plot with initial temperature $t_c < T < t_0$ (where $t_c$ is the temperature at the cusp of the Gibbs energy) is a large black hole.  If the temperature is increased to $t_0$ it becomes thermodynamically favourable for the system to undergo a zeroth order phase transition and jump to the lower branch, corresponding to a small black hole.  If the temperature is further increased to $t_1$, the system will undergo a first order phase transition back to a large black hole.  This is a reentrant phase transition because the thermodynamically preferred state transitions from large to small and then back to large through monotonically increasing the temperature.   

As seen in rightmost plot, above a certain pressure, the branch of the Gibbs energy which terminates due to the entropy condition does so for a temperature less than the temperature at which the Gibbs energy has a cusp.  When this occurs, there is no longer a reentrant phase transition, since the system will never have to ``jump" to be on the minimal branch.  For pressures above this threshold, but less than the critical pressure, we observe only a first order phase transition.  

The situation is captured equally well in Figure~\ref{coexistence_reentrant} which shows the coexistence plots for the $q=e=1$ case.  Here the zeroth order phase transition is marked by the red line, while the black line corresponds to the first order phase transition. We see that the first order phase transition begins at the virtual triple point where the zeroth/first order transitions occur for the same physical parameters, and terminates at the critical point in typical van der Waals fashion.  To the left of the blue line in these plots the black holes have negative entropy.  
   
The analysis which we have here spelled out for the specific case $q=e=1$ applies just as well to any $(q, e)$ combination from the blue region of Figure~\ref{q_e_scrape}.  However, as one moves closer to the zero entropy boundary (the black line in Figure~\ref{q_e_scrape}), the range of pressures for which phase transitions are observed shrinks considerably. 
For example, when $q=1.33$ and $e=1$, the range of pressures for which phase transitions are observed has a width of about $2 \times 10^{-5}$ (for $e=1$, the zero entropy case occurs at $q \approx 1.3375$).
 We deal with the zero entropy case explicitly in a later section.

\subsection{Hyperbolic}

We now move on to consider the hyperbolic case (where $k=-1$).  The uncharged case is uninteresting: regardless of the sign of $q$ there are no physical critical points with positive $(v_c, T_c, P_c)$.  Therefore we move directly to the charged case with non-vanishing $e$.  

There are no critical points in the case $q<0$.  This can be seen easily by considering the equation  of state, which we repeat here for the specific case of $q<0$:
\be 
P_{q <0} = \frac{T}{v} + \frac{2 }{3 \pi v^2} + \frac{512}{243}\frac{e^2}{\pi v^6} + \frac{64}{81}\frac{|q|}{\pi v^5} \, .
\ee
Notice that every term in the equation of state contributes positively to the pressure.  This means that the conditions for a critical point,
\be 
\frac{\partial P}{\partial v} =0, \quad \frac{\partial^2 P}{\partial v^2} = 0 \,, 
\ee
can never be satisfied for real values of the parameters.  Notably, all terms in $\partial^2 P/\partial v^2$ will be positive, and so the equation of state has no inflection points.  Furthermore, there are no phase transitions present at all.  The Gibbs free energy has only a single physical branch, and positivity of entropy is trivially satisfied and no zeroth order phase transitions occur due to this constraint. 

The case $q >0$ is more complicated but not significantly different.  Since $q$ is positive, there will be one negative contribution to the equation of state which allows for the  possibility of critical points.  However, for a given fixed $q>0$, there are threshold values of $e$ given by,
\be 
e_t^{\pm} = \pm  \frac{(50 q^2)^\frac{1}{3}}{8}
\ee
such that whenever $|e| > |e_t^{\pm}|$ the critical temperature is negative---the critical point is unphysical above this threshold.  In addition to requiring $T_c > 0$, 
for $q >0$ we must also be aware of the positivity of entropy condition.  We can study the sign of the entropy for possible critical points satisfying $|e| < |e_t^{\pm}|$ by first evaluating the critical volume at $|e| = |e_t^{\pm}|$.  The result is a complicated radical expression which is approximately,
\be 
v_c^{e_t} \approx 0.9408 q^{1/3} \, .
\ee
The entropy in turn is approximately given by,
\be 
\frac{S_c^{e_t}}{\omega_{3(k)}} = \frac{27}{256}(v_c^{e_t})^3 -\frac{5}{8}q \approx -0.5469 q
\ee
which means, since $q>0$, the entropy is always negative at the point where $T_c$ is equal to zero. 
 Furthermore, since for $|e| < |e_t^{\pm}|$ we have $v_c < v_c^{e_t}$ we then have $S_c < S_c^{e_t}$, and so the critical points inside the region bounded by $e_t^{\pm}$ are unphysical due to positive entropy considerations.   The conclusion of these arguments is that for the hyperbolic case with $q>0$, there are no physical critical points.  We observe neither zeroth nor first order phase transitions for this case.  

\subsection{Black holes with zero entropy}

The presence of the hairy parameter $q$ in the entropy means that it is possible for the black holes of this theory of have zero or negative entropy for finite $r_+$.  Here we consider the special case where the entropy is equal to zero at the critical point.  The interesting feature of this case is that with zero entropy, these black holes have no degrees of freedom. 

To guarantee zero entropy at the critical point, we enforce the constraint,
\be\label{q_crit_constraint} 
q = \frac{27}{160} v_c^3
\ee
To obtain a relationship between $q$ and $e$.  The expression can be obtained analytically, but it is too complicated to be worth quoting; the approximate result is,
\be 
q  \approx 1.3375 |e|^{3/2} \, . 
\ee
The line mapped out by the above expression corresponds to the solid black line in Figure~\ref{q_e_scrape}.  As was discussed in that section, below this threshold we see phase structure analogous to that shown in Figures~\ref{gibbs_reentrant}~and~\ref{coexistence_reentrant}, but for ever narrowing ranges of pressure as the threshold is approached.  To study what happens in the limit that the threshold is actually reached we first expand the equation of state about the critical point in terms of the parameters  $\omega = v/v_c-1$ and $\tau = T/T_c-1$ yielding,
\be\label{eos_smallent} 
\rho = \frac{P}{P_c} -1 =  A(e) \tau - B(e) \omega \tau - C(e) \omega^3 + {\cal O}(\tau \omega^2, \omega^4)
\ee
where $A(e)$ through $C(e)$ denote complicated $e$-dependent coefficients which are positive. In what follows we will not write this $e$ dependence explicitly, but the reader should be aware that it is there.   As an example, for the specific case $e=1$ we have $A = 9/4$, $B= 9/4$ and $C=5/2$.  We will argue that there are no phase transitions when the entropy is zero at the critical point, so let us first be very clear about the logic of our argument.  

In general, we say a phase transition occurs when one branch of the Gibbs free energy expressed in terms of its natural variables $T$ and $P$ becomes thermodynamically favoured over another.  To express the Gibbs free energy in terms of its natural variables, one inverts the equation of state to obtain $v$ as a function of $T$ and $P$.  Since the equation of state is generally a polynomial of some degree $n$ in $v$, there are $n$ separate branches of $v$ and therefore $n$ branches of the Gibbs free energy once the substitution for $v$ is made.  In order for the Gibbs free energy to be physically sensible, then, each of the branches of $v$ must be physically sensible (e.g. $v$ must be positive and must correspond to a black hole with non-negative entropy).  In particular this means that a necessary condition to observe a phase transition is the existence of at minimum two physical branches of the Gibbs free energy---or, equivalently, two physical branches of $v$.  Here we will show that if the entropy is zero at the critical point, then there is only one physically sensible branch for $v$, and hence the system will undergo no phase transitions.  

We can invert  (\ref{eos_smallent}) to obtain three expressions for~$\omega$,
\ba 
\omega_1 &=& \frac{12^{1/3}}{6C} \left[\frac{R(\rho, \tau)^2 - BC 12^{1/3} \tau}{ R(\rho ,\tau)}\right] \, ,
\nn\\
\omega_2 &=& \frac{\sqrt{3} 12^{1/3}}{12 C} \left[\frac{12^{1/3}BC\tau - R(\rho, \tau)^2}{\sqrt{3}R(\rho ,\tau)} \right.
\nn\\
&& \left. +  \frac{12^{1/3} BC \tau + R(\rho, \tau)^2}{R(\rho, \tau)} i\right]\, ,
\nn \\
\omega_3 &=& \omega_2^* \, ,
\ea
where
\be 
R(\rho,\tau) = \left[9AC^2 \tau - 9 \rho + \sqrt{3} C^2 \sqrt{\frac{27(A\tau-\rho)^2 + 4B^3 \tau^3}{C}}\right]^{1/3} 
\ee
Ensuring that at least two solutions for $\omega$ are real amounts to solving the condition,
\be 
12^{1/3} BC \tau + R(\rho, \tau)^2 = 0
\ee
which gives,
\be 
\rho = A\tau - \frac{\left(-12^{1/3}BC \tau \right)^{3/2}}{9 C^2}
\ee
which, for $\tau < 0$ allows us to study the behaviour of the system near the critical point with $T < T_c$ and $P< P_c$.  Using this to write the expressions for the $\omega$'s gives,
\ba
\omega_1 &=& \frac{2}{3} \sqrt{\frac{-3B\tau}{C}}
\nn\\
\omega_2 &=& \omega_3 = -\frac{1}{3} \sqrt{\frac{-3B\tau}{C}} \, .
\ea
Note  $\omega_2$ and $\omega_3$ being negative means $v < v_c$ for those branches, while $\omega_1 > 0$ means $v > v_c$ for that branch.  

Next consider the entropy written in terms of $\omega$ and the constraint~\eqref{q_crit_constraint},
\be 
S_i = \frac{27 \pi^2 v_c^3}{128} \left( (\omega_i+1)^3-1\right)
\ee
since $\omega_2 = \omega_3 < 0$, we have $S_2 = S_3 < 0$ and so these branches are unphysical.  Thus, in the situation where the critical point occurs for zero entropy, the Gibbs free energy has only a single physical branch and therefore there can be no phase transitions. Essentially, if the entropy is zero at the critical point, then it is less than zero for all but one branch of the Gibbs free energy for $T < T_c$ (since the above result assumes $\tau <0$). This result sits well with our thermodynamic intuition, since we would not expect a system possessing zero degrees of freedom to undergo processes like phase transitions.  It is reassuring that this formalism has produced this result.

\section{Conclusions}

We have studied the extended phase space thermodynamics for hairy AdS black hole solutions to Einstein-Maxwell-$\Lambda$ theory conformally coupled to a scalar field in five dimensions.  We have considered spherical, planar, and hyperbolic horizon topologies.  In the planar case, the black hole has no hair and therefore there are no interesting new results.  In the hyperbolic case, physicality constraints on the entropy and critical values eliminate the entire parameter space, yielding no interesting phase structure.  

The spherical case has yielded interesting results. These  systems show van der Waals behaviour in the charged and uncharged cases.  In the zero entropy limit (which occurs for non-zero $r_+$ for positive $q$), all critical behaviour ceases.  In the charged case, demanding the positivity of entropy led us to find examples of reentrant phase transitions for these solutions. This is the first example of a reentrant phase transition in a five dimensional gravitational system which does not involve higher curvature corrections.

\section*{Acknowledgments}
This research was supported in part by the Natural Sciences and Engineering Research Council of Canada through their discovery and PGS programmes.

\appendix
\section{Black holes with zero or negative mass}

As mentioned in the body of the paper, the metric~\eqref{metric} admits zero and negative mass solutions for $k = \pm 1$ for $q >0$.  These negative mass solutions were regarded as pathological in~\cite{Galante:2015voa, Giribet:2014fla}, and so they were not studied in detail.  However, the idea of negative mass black holes is not new: negative mass solutions for $k=-1$ have been known for quite some time as solutions to Einstein-AdS gravity, and can be formed by the collapse of matter with negative energy density \cite{Smith:1997wx, Mann:1997jb}.  For these reasons, this topic deserves a bit more discussion here.

First we consider the spherical case (i.e. $k=+1$) where we find that, provided $q >0$, there can be massless or negative mass black holes.  For vanishing electric charge, there is a single horizon provided $q > 0$.  When $e \not= 0$, for a given $m_0 \leq 0$, there is a minimum value of $q_0$ required for horizons to exist which is given by,
\be 
q_0 = \frac{2\left(e^2l^2 - 2r_0^6 -l^2 r_0^4 \right)}{l^2 r_0}
\ee
where $r_0$ solves the equation,
\be 
m_0 = \frac{5r_0^6 + 3l^2r_0^4 -e^2l^2}{r_0^2l^2}
\ee
For $q > q_0$ the solution has an inner/outer horizon structure, while for $q < q_0$ the solution describes a naked singularity.  When $q=q_0$, the two horizons coincide and the solution is extremal.  

 It is fairly straightforward to show that in the $k=+1$ case  the negative mass black holes also have negative entropy. Written in terms of the event horizon radius, the mass parameter reads,
\be 
m = \frac{r_+^6 + l^2(kr_+^4 +e^2 - qr_+)}{r_+^2 l^2} \, .
\ee
A negative mass solution results when the numerator of this expression is less than zero.  Inverting the expression $s=S/\omega_{3(k)}$ for $r_+$, the condition for a negative mass black hole becomes,
\ba 
0 &>&  \frac{5l^2\left(20 q + 32 s \right)^{1/3}}{4}\left[\left( k - \frac{2}{5}\right)q + \frac{8k}{5}s  \right] 
\nn\\
&& +\frac{25}{4}\left(q +\frac{8}{5}s \right)^2 + e^2 l^2 
\ea
which, for $k=+1$, can  be satisfied only for $s <0 $ since $q >0$.  Therefore, in the spherical case, the inclusion of these negative mass solutions adds nothing further to our analysis, since we have enforced positivity of entropy throughout our work.

The situation is significantly more interesting in the hyperbolic case (i.e. $k=-1$).  We start by studying the negative mass solutions for the case where $ e= 0$.  For this case it is not necessary to have $q$ strictly positive to have negative mass black hole solutions.  For $0 > m_0 > -l^2/4$. there will be black hole solutions with and $q<0$  provided $q > q_0$ where,
\be 
q_0 = \frac{2r_0^3}{l^2}\left(l^2 - 2r_0^2 \right)
\ee
with
\be 
r_0^2 = \frac{1}{10} \left(l\sqrt{9l^2 + 20 m_0}  + 3l^2  \right) \, .
\ee
So long as $0 > q > q_0$ the black hole exhibits an inner/outer horizon structure, with the two horizons coinciding when $q = q_0$ corresponding to an extremal black hole.  For $ q < q_0$ there is a naked singularity. 

When $q > 0$ the negative mass black holes can have up to three distinct horizons for a small range of parameters. This occurs for $ 0\geq  m \geq -9 l^2/20$, and for a given $m_0$ in this interval takes place for $q_{-} \leq q \leq q_{+}$ with
\be 
q_{\pm}^2  = \frac{2r_{0_\pm}^3}{l^2}\left(l^2 - 2r_{0_\pm}^2 \right)
\ee
where
\be 
r_{0_\pm}^2 = \frac{1}{10} \left(3l^2 \pm l\sqrt{9l^2 + 20 m_0}    \right) \, .
\ee
Outside of this range, the solution has a single horizon.  

With three horizons, there are a number of interesting possibilities.  For example, one could consider three different `extremal limits' corresponding to when the inner and intermediate horizons coalesce ($q=q_+$), the outer and intermediate horizons coalesce ($q=q_-$), or when all three horizons coalesce ($q=q_-=q_+$). Let us consider explicitly the geometry of this third case.

The triple coincidence of horizons corresponds to a triple root of the metric function, with the specific parameters,
\be 
m = -\frac{9}{20}l ^2\, \quad q = \frac{3\sqrt{30}}{125} l^3\,\quad r_+ = \frac{\sqrt{30}}{10} l \, .
\ee

 The solution in this limit is in fact free from curvature singularities at the horizon: the Kretschmann scalar is given by
\be 
K = \frac{400}{3 l^4} \, .
\ee  
To study the near horizon geometry we expand
\be 
f(r) \approx \frac{1}{6} f'''(r_+)(r-r_+)^3
\ee
giving for the Euclidean section of the metric ($t \to i\tau$),
\be 
ds^2 \approx \frac{16}{9 (f'''(r_+))^2 R^6} d\tau^2 + dR^2 + r^2(R)d\Sigma^2_3
\ee
where 
\be 
dR^2 = \frac{6dr^2}{f'''(r_+)(r-r_+)^3} \, .
\ee
and we are interested in the large $R$ limit (corresponding to $r \to r_+$). 
On the $(\tau, R)$ section of the spacetime, the Ricci scalar reads,
\be 
\sccur = - \frac{24}{R^2} \, .
\ee
This is a space of non-constant negative curvature, which is singular at $R=0$, corresponding to $(r-r_+) \to \infty$.  In our expansion of $f$ above, it was assumed that we were looking at the metric close to the event horizon.  Clearly, $(r-r_+) \to \infty$ violates this assumption, meaning the curvature singularity is not a physical pathology of the spacetime. 

We can gain further insight into this geometry by, near the horizon, writing the metric function as, 
\be 
f(r) \approx \frac{1}{6} f'''(r_+)(r-r_+)^3  = a(r-r_+)^3
\ee
so we can write the Euclidean $(t,r)$ section as,
\be 
ds^2 = a(r(\tilde{r})-r_+)^3 \left[dt^2 + d\tilde{r}^2 \right]
\ee
with
\be 
 d\tilde{r}^2 = \frac{dr^2}{ a^2(r-r_+)^6} \, .
\ee
Transforming to 
\be 
\rho = e^{\beta \tilde{r}}, \quad \tau = \beta t 
\ee
the metric reads
\be 
ds^2= \frac{a(r(\rho)-r_+)^3}{\beta ^2 \rho^2} \left[\rho^2 d\tau^2 + d\rho^2 \right] \, .
\ee
However, because here we have
\be 
\tilde{r} = \frac{1}{2a(r-r_+)^2}\,,
\ee
$\rho$ has an essential singularity at $r=r_+$. This means that no matter how $\beta$ is chosen, the conformal factor cannot be made regular at the horizon, and so there is no issue of a conical singularity there.

Since there is no issue of a conical singularity, the derivation of the temperature by Euclidean methods fails.  However, we can calculate the surface gravity at the horizon and find it to be,
\be\label{surface_gravity} 
\kappa = \frac{f'(r_+)}{2} \, .
\ee
Since we are considering a multiple root of $f$, the derivative vanishes at the horizon and the surface gravity is $\kappa = 0$.  Therefore these extremal black holes have zero temperature.

In the case of nonzero electric charge the situation is quite similar except now there is the possibility of up to four horizons.  Like the uncharged case, we can consider limits where various combinations of these horizons coincide.  The inclusion of charge results in a polynomial complicated enough that it is hard to, in a meaningful way, express the ranges over which a certain number of horizons exist.  For this reason here we focus on the solution when four horizons coincide and examine its geometry.

Four horizons will coincide provided,
\ba 
m &=& -\frac{3}{5}l^2\,, \quad q = \frac{16\sqrt{5}}{125}l^3\,, 
\nn\\
e &=& \pm \frac{l^2}{5}\, ,\quad  r_+ = \frac{l}{\sqrt{5}}\, .
\ea
The Kretschmann scalar in this case is given by, $K~=~300/l^4$, which is obviously well-behaved everywhere.  
For this solution, the metric function can be expanded near the horizon as,
\be 
f(r) \approx \frac{f^{(4)}(r_+)}{4!}(r-r_+)^4 = \frac{50}{l^4}(r-r_+)^4
\ee
The Euclidean ($t,r)$ section of the metric then can be written as,
\be 
ds^2 \approx \frac{50(r-r_+)^4}{l^4 \beta^2 \rho^2} \left[\rho^2 d\tau^2 + d\rho^2 \right]
\ee
where 
\be 
\rho = \exp\left(\frac{\beta l^4}{50 (r-r_+)^3} \right)\,, \quad \tau = i\beta t\, .
\ee
As before, $\rho$ has an essential singularity at the horizon, and so the conformal factor cannot be made regular there and there are no conical singularities.   The surface gravity, given by eq.~\eqref{surface_gravity}, vanishes and the temperature is zero.

As we have seen, the hyperbolic negative mass black holes exhibit a rich and interesting structure with up to four horizons.  While it is true that the negative mass solutions add nothing new to our thermodynamic considerations in the main text (since the spherical ones have negative entropy, and we showed earlier that there are no interesting phase behaviour for the hyperbolic black holes), these exotic possibilities present new and interesting behaviour.  It would be worthwhile to better understand the causal structure of these black hole geometries.

\providecommand{\href}[2]{#2}\begingroup\raggedright\endgroup


\end{document}